\documentclass[aps,prl,reprint,amsmath,amssymb,superscriptaddress]{revtex4-2}

\usepackage{amsmath}
\usepackage{graphicx}
\usepackage{dcolumn}
\usepackage{bm}
\usepackage{multirow}
\usepackage{subfigure}
\usepackage{booktabs}
\usepackage{hhline}  
\usepackage{float}
\usepackage{braket}
\usepackage[colorlinks,linkcolor=blue,anchorcolor=blue,citecolor=blue,urlcolor=blue]{hyperref}
\usepackage[normalem]{ulem}

\begin{document}

\title{Enhancing superconducting transition temperature of lanthanum superhydride by increasing hydrogen vacancy concentration}


\author{Haoran Chen}
\email{chenhr1115@pku.edu.cn}
\affiliation{International Center for Quantum Materials, Peking University, Beijing 100871, China}

\author{Hui Wang}
\affiliation{Key Laboratory for Photonic and Electronic Bandgap Materials (Ministry of Education), School of Physics and Electronic Engineering, Harbin Normal University, Harbin 150025, China}

\author{Junren Shi}
\affiliation{International Center for Quantum Materials, Peking University, Beijing 100871, China}
\affiliation{Collaborative Innovation Center of Quantum Matter, Beijing 100871, China}


\date{\today}
	
\begin{abstract}
	Various clathrate superhydride superconductors have been found to possess hydrogen deficiency in experimental samples, while their impacts on superconductivity are often neglected. In this study, we investigate the superconductivity of lanthanum superhydride with hydrogen deficiency (LaH$_{10-\delta}$) from first principles using path-integral approaches.
	Under the effects of thermal and quantum fluctuations, 
	hydrogen vacancies are found to diffuse within the system, leading to modifications in ion vibrations, electronic structure and electron-phonon coupling.
	These changes result in a non-monotonic dependence of superconducting transition temperature ($T_c$) on the vacancy concentration ($\delta$).
	By comparing the experimental and theoretical equations of state, we suggest that $\delta$ varies across samples under different pressures. This explains the positive pressure dependence of $T_c$ in experiments below 150~GPa.
	Remarkably, within this pressure range, we find that $T_c$ could be further raised by increasing $\delta$.
\end{abstract}


\maketitle



\textit{Introduction.}—
Superconductivity, ever since its discovery in 1911, has been a subject of sustained interest in the area of physics.
The continuous quest to achieve higher superconducting transition temperatures ($T_c$'s) has driven extensive research efforts.
In conventional superconductors, phonons and their coupling with electrons are identified as crucial factors. As pointed out by the celebrated Bardeen-Cooper-Schrieffer (BCS) theory, raising phonon frequencies and enhancing electron-phonon coupling (EPC) directly lead to higher $T_c$'s~\cite{bardeen_theory_1957,eliashberg_interactions_1960,mcmillan_transition_1968,allen_transition_1975}.
Over the past decade, the idea has guided the discovery of a large family of hydride superconductors~\cite{ashcroft_metallic_1968,ashcroft_hydrogen_2004,duan_pressure-induced_2014,drozdov_conventional_2015,liu_potential_2017,drozdov_superconductivity_2019,kong_superconductivity_2021,ma_high-temperature_2022,wang_superconductive_2012,li_superconductivity_2022,di_cataldo_mathrmmathrmbh_8_2021,liang_prediction_2021,zhang_design_2022,song_stoichiometric_2023}. Because of the light mass of hydrogen ions, the systems can exhibit high phonon frequencies and thus high $T_c$'s.
Beyond conventional superconductors, more other properties, such as  electron correlation strength or system dimension, could significantly affect superconductivity.

In real materials, the quest to enhance $T_c$ thus often involves exploring new approaches to modify these properties.
For example, externally, applying pressure is a commonly used technique. It cannot only directly tune various properties such as phonon spectrum, EPC and electronic structure~\cite{tomita_dependence_2001,kasinathan_superconductivity_2006,profeta_superconductivity_2006,bazhirov_superconductivity_2010}, but can also enable the stabilization of various superconducting phases that are unstable under atmosphere conditions.
This approach has notably contributed to the discovery of hydride superconductors~\cite{ashcroft_metallic_1968,ashcroft_hydrogen_2004}.
Internally, we can incorporate new chemical elements into the system. 
These new elements help explore the structural phase space with high $T_c$'s as in hydride superconductors~\cite{sun_route_2019,liang_prediction_2021,zhang_design_2022,semenok_superconductivity_2021,bi_giant_2022,chen_enhancement_2023,huang_synthesis_2023,chen_superior_2024,zhang_high-pressure_2024}, or dope electrons and holes as in unconventional ones.
In some cases, superconductivity could even be enhanced by introducing impurities or inhomogeneity~\cite{martin_enhancement_2005,gastiasoro_enhancing_2018,slebarski_superconductivity_2014,slebarski_enhancing_2020}, although a first-principles understanding of these effects is still lacking.

In clathrate hydride superconductors, manipulation of hydrogen vacancies could present a potential new approach for controlling superconductivity. 
Recent studies have revealed that these vacancies could strongly affect the properties of the hydrides~\cite{wang_quantum_2021,chen_coexistence_2024}, including ion dynamics, electronic structure, EPC and superconducting properties.
With the assistance of thermal and quantum fluctuations, hydrogen ions could even diffuse between different vacant sites, rendering the system a superionic superconductor. 
On the other hand, such vacancies are not uncommon. They have been observed in several different clathrate superhydrides, such as intrinsically in Li$_2$MgH$_{16}$~\cite{wang_quantum_2021,chen_coexistence_2024}, or as deficiencies in LaH$_{10}$, YH$_{6/9}$ and CaH$_{6}$~\cite{drozdov_superconductivity_2019,wang_quantum_2023,kong_superconductivity_2021,ma_high-temperature_2022,wang_unveiling_2023}. 
By manipulating these vacancies, it may be possible to fine-tune and enhance the superconductivity in these systems.

To explore the possibility, in this Letter, we study the superconductivity in lanthanum superhydride with hydrogen deficiency (LaH$_{10-\delta}$). 
Experimental samples of this material have been estimated to possess hydrogen vacancies of about 4\%~\cite{drozdov_superconductivity_2019}, with evidence of these vacancies diffusing in the system~\cite{wang_quantum_2023}.
We investigate the dependence of $T_c$ on both pressure and vacancy concentration ($\delta$) using first-principles calculations.
In the presence of anharmonic vibrations, quantum fluctuations and ion diffusion, conventional harmonic theories become inapplicable.
To correctly describe these effects, we apply the non-perturbative stochastic path-integral approach (SPIA), which directly analyzes the pairing between electrons induced by moving ions from path-integral molecular dynamics (PIMD) simulations~\cite{liu_superconducting_2020,chen_first-principles_2021,chen_stochastic_2022,zhang_nonperturbative_2022,chen_coexistence_2024}.
Our results reveal a non-monotonic dependence of $T_c$ on $\delta$. 
By comparing experimental and theoretical equations of state, we determine the position of experimental samples on the $T_c$\,-\,$\delta$ curve, demonstrating that at 137~GPa, $T_c$ could be further raised by increasing hydrogen vacancies.
%
Additionally, the deviation of $\delta$ at 137~GPa from higher-$T_c$ values leads to a positive pressure dependence of $T_c$ between 137 and 150~GPa, consistent with previous experimental observations.


\begin{figure}[t]
	\includegraphics[width=8.6cm]{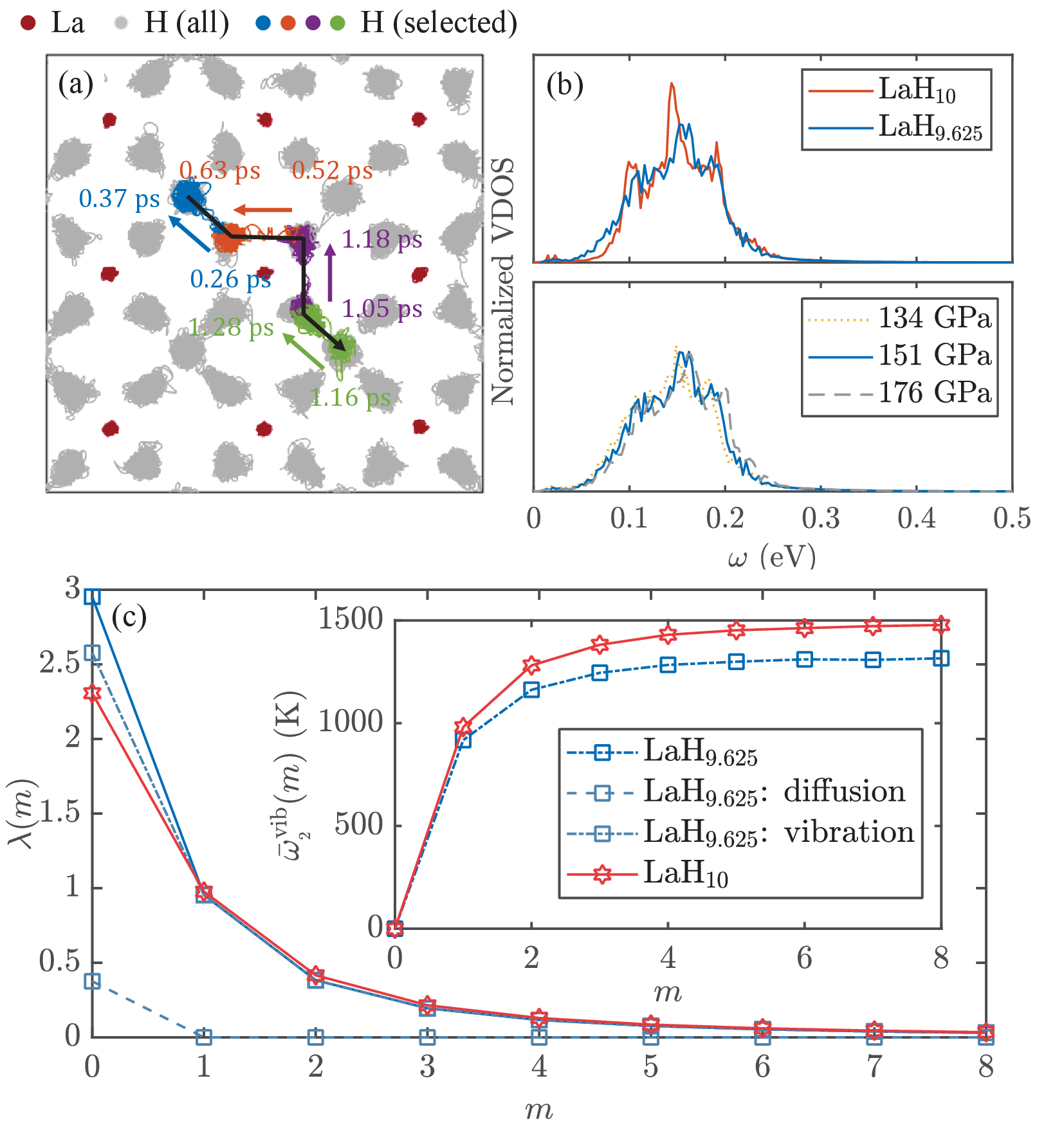}
	\caption{\label{fig:trajectories} 
		(a) [001] view of trajectories of the centroid mode of all hydrogen (grey) and lanthanum (red) ions in  LaH$_{9.625}$ under 151~GPa from PIMD simulations at 240~K. 
		Trajectories of four selected hydrogen ions are highlighted with different colors, as well as the directions, approximate start time and end time of diffusion.
		Effectively, the hydrogen-vacancies diffuse in the opposite direction to these ions, as schematically illustrated by the black arrow.
		(b) Top: vibrational density of states (VDOS) calculated in LaH$_{9.625}$ under 151~GPa and LaH$_{10}$ under 158~GPa. Bottom: VDOS of LaH$_{9.625}$ under different pressures. The VDOS's are calculated from the Fourier transformation of velocity autocorrelation function of the centroid mode in ring-polymer molecular dynamics simulations. 
		(c) EPC parameters $\lambda(m)$ of LaH$_{9.625}$ (LaH$_{10}$) under 151~GPa (158~GPa) at 240~K.
		For LaH$_{9.625}$, we also show the diffusional contribution ($\lambda^{\mathrm{dif}}$) and vibrational contribution ($\lambda^{\mathrm{vib}}$), respectively.
		Inset: the asymptotic behavior of $\bar{\omega}_2^{\mathrm{vib}}(m)$ in Eq.~(\ref{omega2}). The zero-frequency contribution from diffusion is not taken into account.
	}
\end{figure}

\textit{Diffusing vacancies and superconductivity.}—
To begin, we first focus on a specific chemical stoichiometry LaH$_{9.625}$~\cite{drozdov_superconductivity_2019,wang_quantum_2023} and study how the properties of the system are affected by hydrogen vacancies.
We perform PIMD simulations at 240~K, which is close to the experimental $T_c$, to describe the defective system~\cite{marx_ab_1996,ceriotti_efficient_2010}.
Since some hydrogen ions are removed from the system, it becomes frequently the case that hydrogen ions neighbor a vacant site, making it easy for ions to diffuse or hop between different sites with the assistance of quantum and thermal fluctuations~\cite{wang_quantum_2023,wang_quantum_2021,chen_coexistence_2024}.
In Fig.~\ref{fig:trajectories}~(a), we show the trajectories of centroid-mode ions in PIMD simulations. 
It can be seen that some hydrogen ions become diffusive. 
This behavior is similar to the superionic phase observed in the stoichiometric LaH$_{10}$, in which hydrogen ions begin to diffuse at temperatures above 800~K~\cite{liu_dynamics_2018,causse_superionicity_2023}.
However, unlike in LaH$_{10}$, where all hydrogen ions participate in diffusion, in LaH$_{9.625}$, diffusion is primarily associated with the vacancies.
This is illustrated by the diffusion paths of several selected hydrogen ions in Fig.~\ref{fig:trajectories}~(a). 
Individually, all of the selected ions stop diffusing after occupying a previously vacant site, while their diffusion paths together form a longer path, along which the vacancy diffuses in the inverse direction.

For hydrogen ions not on the diffusion path, they still vibrate around their ``equilibrium positions'' like solids.
Compared with LaH$_{10}$, 
the vibration is softened (see Fig.~\ref{fig:trajectories}~(b)). This is similar to liquids, in which the low-frequency vibrational density of states is broadened by ion diffusion~\cite{jaffe_superconductivity_1981}.
On the other hand, the ``equilibrium structure'' is distorted from that of LaH$_{10}$ by vacancies. Since the positions of the vacancies vary over time, the distortion also dynamically evolves. 


The unusual ion motion above obviously break the validity of conventional harmonic theories in solids. It can also be expected that the electronic structure will be modified in this dynamically distorted system~\cite{noauthor_see_nodate}. To non-perturbatively take into these effects, we apply the SPIA method to determine the normal-state electron states and the pairing between them ~\cite{liu_superconducting_2020,chen_first-principles_2021,chen_stochastic_2022,zhang_nonperturbative_2022,chen_coexistence_2024,noauthor_see_nodate}.
In Fig.~\ref{fig:trajectories}~(c), we compare the the EPC parameters of LaH$_{10}$ and LaH$_{9.625}$.
The defective system has a static EPC parameter $\lambda(0)=2.95$, stronger than that of the perfect lattice, in which $\lambda(0)=2.3$
\footnote{$\lambda(0)$ of LaH$_{10}$ in our calculations is smaller than previous results obtained by SSCHA. The deviation could originate from either the way of dealing with anharmonic effects, or the finite size effect in our simulations, in which supercells containing 352~atoms are used.}. 
For higher-frequency components, on the other hand, the values are slightly suppressed in the defective lattice.
Using a typical Morel-Anderson pseudopotential $\mu^{*}=0.10$ to account for the effect of Coulomb repulsion interaction~\cite{morel_calculation_1962}, we solve the linearized Eliashberg equations. We find that $T_c$ of LaH$_{9.625}$ is 236~K, which is slightly suppressed from 246~K of LaH$_{10}$.
Both values are close to the experimental measurements and previous theoretical results~\cite{drozdov_superconductivity_2019,errea_quantum_2020}.

The observation that LaH$_{9.625}$ exhibits a larger $\lambda(0)$ but a lower $T_c$ is a characteristic of diffusive systems~\cite{liu_superconducting_2020,chen_first-principles_2021,chen_coexistence_2024}.
As discussed in our previous work~\cite{chen_coexistence_2024}, the coupling of electrons with the ion diffusion is responsible for the enhancement of $\lambda(0)$. 
On the other hand, the diffusion barely contributes to higher-frequency components, which are mainly related with local vibrations around the diffusion paths.
By solving the Eliashberg equation, it can be found that it is these finite-frequency EPC parameters that determine $T_c$. As a result, the larger $\lambda(0)$ does not lead to a higher $T_c$.
This is also the case for LaH$_{10-\delta}$.
By subtracting the contribution from ion diffusion, we find that the vibrational part $\lambda^{\mathrm{vib}}(0)\approx2.55$ of LaH$_{9.625}$ is close to that of LaH$_{10}$.
On the other hand, we can infer the EPC-weighted average phonon frequency from the asymptotic behavior of high-frequency components of EPC parameters~\cite{allen_superconductivity_1972}, i.e.,
\begin{eqnarray}\label{omega2}
	\bar{\omega}_2^{\mathrm{vib}}
	=2\pi/\hbar\beta\left.\sqrt{m^2\lambda^{\mathrm{vib}}(m)/\lambda^{\mathrm{vib}}(0)}\right|_{m\rightarrow\infty}.
\end{eqnarray}
It can then be seen that the softening of vibrational spectral (see Fig.~\ref{fig:trajectories}~(b)) leads to a smaller EPC-weighted average phonon frequency $\bar{\omega}_2^{\mathrm{vib}}$ (about 1490~K in LaH$_{10}$ and 
about 1325~K in LaH$_{9.625}$), resulting in the slight suppression of $T_c$.

\begin{figure}[t]
	\includegraphics[width=8.6cm]{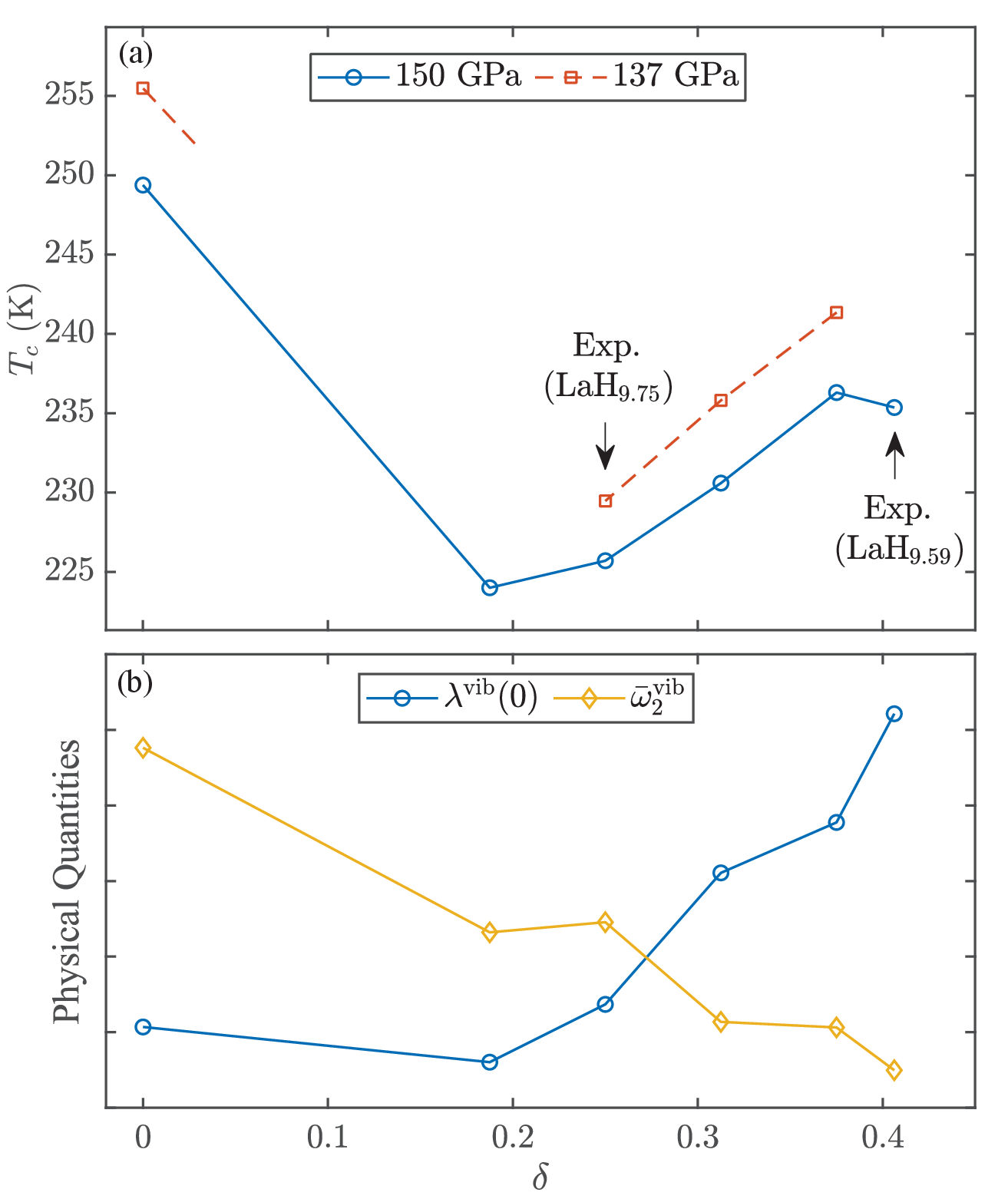}
	\caption{\label{fig:Tc_con} (a) vacancy-concentration dependence of superconducting $T_c$ of LaH$_{10-\delta}$ at 150~GPa and 137~GPa. The values are obtained by linearly interpolating $T_c$'s at two cloest pressures~\cite{noauthor_see_nodate}.
	(b)	vacancy-concentration dependence of other physical properties around 150~GPa, including the vibrational part of static EPC parameter $\lambda^{\mathrm{vib}}(0)$ (blue circles, 2.2\,-\,2.8) and EPC-weighted average phonon frequency $\bar{\omega}_2^{\mathrm{vib}}$ (yellow diamonds, 1275\,-\,1550~K). 
	These results are from simulations in supercells of volume 1062.39~\AA$^3$, containing 32 formulas of LaH$_{10-\delta}$. 
	}
\end{figure}

\textit{Enhance $T_c$ by increasing vacancy concentration.}—
From the discussions above, we see that the introduction of hydrogen vacancies greatly changes the ion motion of the system, leading to the modification of EPC and superconductivity.
On the other hand, the concentration of vacancies determines the number of vacant sites available for hydrogen ion diffusion, as well as the number of electrons in the system.
As a result, it can be expected that superconductivity will be modified upon the change of vacancy concentration.
In Fig.~\ref{fig:Tc_con}~(a), we adjust $\delta$ in LaH$_{10-\delta}$ and calculate the corresponding $T_c$ under fixed external pressures.
The results reveal that as the concentration of vacancies increases, the predicted $T_c$ initially decreases and then subsequently rises.

\begin{figure}[t]
	\includegraphics[width=8.6cm]{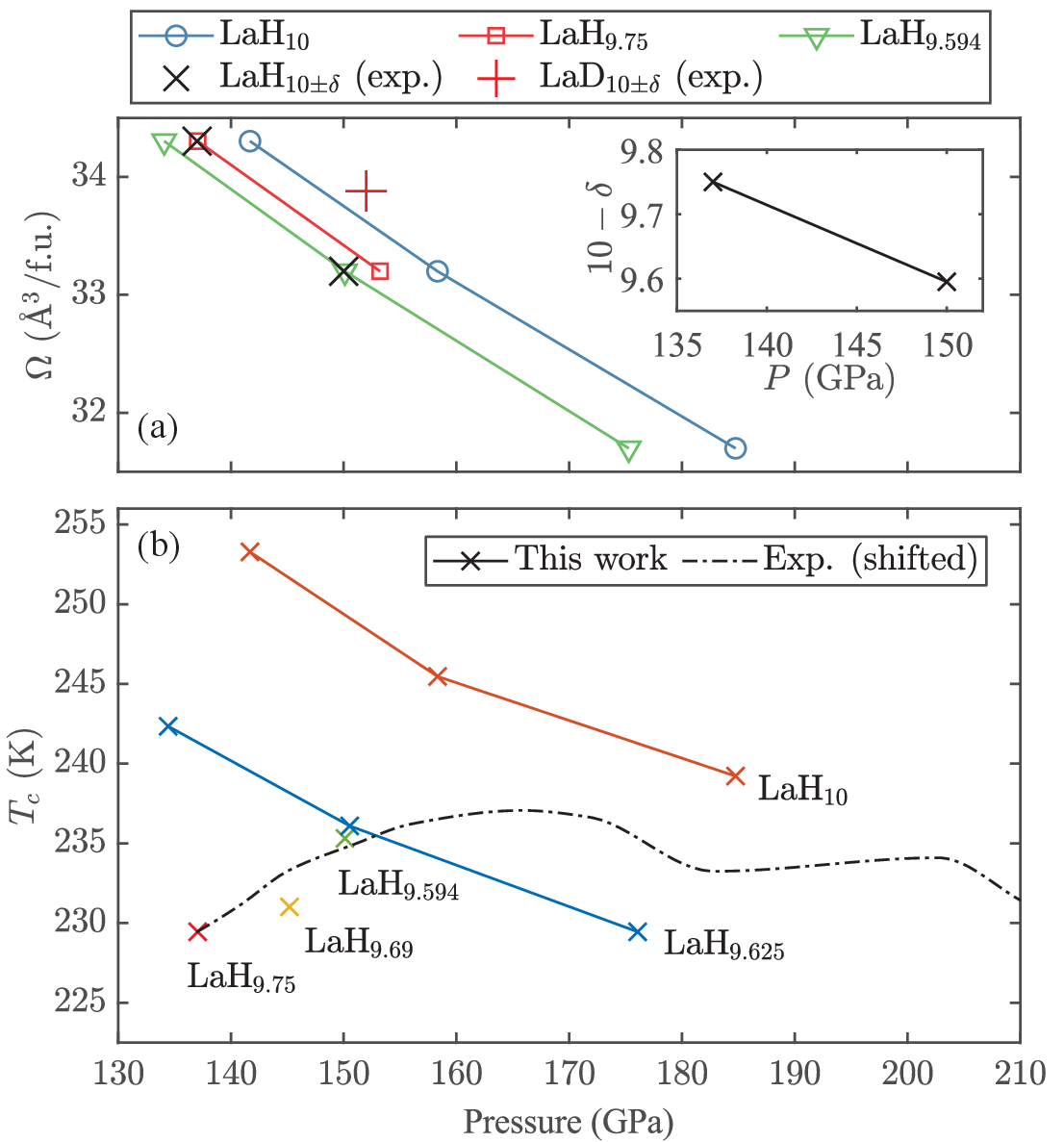}
	\caption{\label{fig:Tc_EOS}
		(a) Equation of states of LaH$_{10-\delta}$ with different $\delta$'s calculated from PIMD simulations under 240~K.
		Experimental results from Ref.~\cite{drozdov_superconductivity_2019} are shown for comparison. 
		Inset: experimental hydrogen stoichiometry $10-\delta$ estimated from the equation of states.
		(b) Pressure dependence of superconducting $T_c$ of LaH$_{10-\delta}$ with different vacancy-concentrations. 
		Experimental measurements (downshifted by 13~K) from Ref.~\cite{drozdov_superconductivity_2019} are shown for comparison.
		More results can be found in the Supplementary Materials~\cite{noauthor_see_nodate}.
	}
\end{figure}

Does the observation above indicate the possibility of enhancing $T_c$ through the adjustment of vacancy concentrations in experiments? 
To examine the possibility, it is essential to determine the position of experimental samples on the $T_c$\,-\,$\delta$ curve. 
In Fig.~\ref{fig:Tc_EOS}~(a), we follow Ref.~\cite{wang_quantum_2023} and compare the equation of states of LaH$_{10}$ from experiments and PIMD simulations, in which the quantum pressure is estimated using the centroid-virial estimator introduced in Ref.~\cite{ceriotti_i-pi_2014}. 
An obvious discrepancy is observed, indicating that there are less hydrogen than expected in stoichiometric LaH$_{10}$. 
By changing $\delta$ in PIMD simulations, we find that the experimental samples have hydrogen vacancies of approximately 2.5\% (LaH$_{9.75}$) at 137~GPa and 4.1\% (LaH$_{9.59}$) at 150~GPa.
These observations are consistent with experimental estimations and previous theoretical studies~\cite{drozdov_superconductivity_2019,wang_quantum_2023}. 

Substituting the experimental $\delta$ values into Fig.~\ref{fig:Tc_con}~(a), it can be seen that, at 137~GPa, the vacancy concentration of the experimentally synthesized sample is within the range where $T_c$ increases.
Increasing $\delta$ from 0.25 to 0.375 enhances $T_c$ by about 12~K.
Conversely, at 150~GPa, the vacancy concentration is close to the locally optimal value. 
%
On the other hand, experimental measurements for other clathrate superhydrides, such as CaH$_{6}$, have shown variations in vacancy concentrations between samples under the same pressure~\cite{ma_high-temperature_2022,wang_unveiling_2023}. This variability suggests that the value of $\delta$ is not an intrinsic property of these systems, and is thus manipulable.
Based on these observations, we predict that $T_c$ of LaH$_{10-\delta}$ at 137~GPa could be enhanced by increasing the vacancy concentration in experiments.


It is then natural to ask what underlies the non-monotonic $T_c$\,-\,$\delta$ dependence.
In conventional theories, one usually uses the zero-frequency EPC parameter $\lambda(0)$ and the average phonon frequency $\bar{\omega}$ to characterize a superconducting system.
In diffusive systems, as discussed above for LaH$_{9.625}$, the diffusion part enhances $\lambda(0)$ but barely contribute to superconductivity. 
By subtracting the diffusion part, the residual $\lambda^{\mathrm{vib}}(m)$ from local vibrations can be considered to be similar to that of solids.
In Fig.~\ref{fig:Tc_con}~(b), we show the dependence of $\lambda^{\mathrm{vib}}(0)$ and EPC-weighted average phonon frequency $\bar{\omega}_2^{\mathrm{vib}}$ on the vacancy concentration.
We see that $\lambda^{\mathrm{vib}}(0)$ behaves similarly to $T_c$. However, systems of higher vacancy concentrations with larger $\lambda^{\mathrm{vib}}(0)$ does not exhibit stronger superconductivity compared with the nearly stoichiometric systems.
This originates from the softening of local vibrations.
As shown in Fig.~\ref{fig:Tc_con}~(b), the average phonon frequencies $\bar{\omega}_2^{\mathrm{vib}}$ keep decreasing with respect to the increasing vacancies, reshaping the $T_c$\,-\,$\delta$ curve to the observed non-monotonic form.

\textit{Experimental pressure-dependence of $T_c$.}—
Lastly, we would like to discuss the relation of the $T_c$\,-\,$\delta$ dependence with existing experimental measurements, focusing specifically on the dependence of $T_c$ on external pressure.
As shown in Fig.~\ref{fig:Tc_EOS}, when the vacancy concentration $\delta$ is held constant, $T_c$ monotonically decreases with increasing pressure.
However, this monotonic behavior is changed when we consider the variation of $\delta$ across samples under different pressures.
Taking LaH$_{9.75}$ at 137~GPa and LaH$_{9.59}$ at 150~GPa, as estimated from the experimental equation of state, we observe an increase of $T_c$ upon compression within this range of pressure.
This trend is consistent with the experimental measurements by Drozdov \textit{et al.}~\cite{drozdov_superconductivity_2019}.

In conventional superconductors, such a non-monotonic $dT_c/dP$ behavior is uncommon~\cite{lorenz_high_2005}. 
Typically, compressing these systems results in a stiffening of vibrational spectrum and a decrease of electronic density of states at Fermi level, which are also observed in LaH$_{10-\delta}$ when $\delta$ is fixed (see Figs.~\ref{fig:trajectories}(b),~\ref{fig:Tc_EOS}(b) and S2). 
According to the well-known McMillan formula, these changes suggest that $T_c$ will decrease monotonically~\cite{griessen_pressure_1987,tomita_dependence_2001}.
A positive dependence of $T_c$ on pressure can indicate the presence of additional factors favorable to superconductivity~\cite{chu_study_1970,chu_direct_1974,chu_pressure-enhanced_1974,kasinathan_superconductivity_2006,profeta_superconductivity_2006,bazhirov_superconductivity_2010}.
In the case of LaH$_{10-\delta}$, our study suggests it is the variation of vacancy concentration that plays the role. 
At 137~GPa, $\delta$ deviates from the higher-$T_c$ value at 150~GPa, leading to a positive $dT_c/dP$ in this region.
Interestingly, in experiments, several other samples below 150~GPa also exhibit lower $T_c$ than that at 150~GPa~\cite{drozdov_superconductivity_2019,noauthor_see_nodate}. As a result, we predict that $T_c$ could also be enhanced in these experimental samples by increasing vacancy concentrations .

\textit{Summary.}— 
In summary, we systematically study the effects of hydrogen vacancies on the properties of LaH$_{10-\delta}$, including ion motion, EPC and superconductivity from first principles.
Through PIMD simulations, we find that the vacancies diffuse in the system with the assistance of thermal and quantum fluctuations.
This ion diffusion softens the ion vibrations and modifies the electronic structures, thus modifying the superconductivity.
By analyzing the contribution to superconductivity from different forms of ion motion, we find that ion diffusion contributions a large static EPC parameter $\lambda(0)$, while the superconductivity is mainly induced by local vibrations of ions around their diffusion paths or equilibrium positions.
By adjusting the vacancy concentration, we observe that within the range of $0.1875<\delta<0.375$ (corresponding to $9.625<10-\delta<9.8125$), increasing vacancies leads to a continuous softening of local vibrations and enhancement of EPC strength $\lambda^\mathrm{vib}(0)$. The interplay of the two factors leads to the increase of $T_c$ when more hydrogen vacancies are introduced.
By comparing the equation of state from experiments and PIMD simulations, we determine the experimental chemical stoichiometry.
At 137~GPa, the stoichiometry of the experimental sample is LaH$_{9.75}$, whose vacancy concentration falls in the middle of the range, suggesting that superconductivity could be further enhanced by increasing hydrogen vacancies.
On the other hand, at 150~GPa, the stoichiometry is estimated to be LaH$_{9.59}$, close to the locally optimal value. 
This variation of $\delta$ across samples leads to a lower $T_c$ at the lower pressure, which is consistent with previous experimental measurements~\cite{drozdov_superconductivity_2019}. 
Such a positive pressure dependence of $T_c$ has also been observed in several other samples below 150~GPa in experiments~\cite{drozdov_superconductivity_2019,noauthor_see_nodate}, suggesting that their $T_c$ might also be raised by increasing vacancies.
%
More generally, the presence of hydrogen vacancies has also been observed in various clathrate superhydrides~\cite{wang_quantum_2021,chen_coexistence_2024,drozdov_superconductivity_2019,wang_quantum_2023,kong_superconductivity_2021,ma_high-temperature_2022,wang_unveiling_2023}.
As a result, the present work presents a potential method to enhance $T_c$ in the family of systems, as well as a possible explanation for the positive pressure-dependence of $T_c$ in such systems~\cite{drozdov_superconductivity_2019,kong_superconductivity_2021,ma_high-temperature_2022,wang_unveiling_2023}.

\nocite{kresse_efficient_1996,kresse_ultrasoft_1999,blochl_projector_1994,perdew_generalized_1996,chen_coexistence_2024,jinnouchi_--fly_2019,jinnouchi_phase_2019,ceriotti_efficient_2010,chen_first-principles_2021,chen_stochastic_2022,liu_superconducting_2020,allen_recovering_2013,zacharias_fully_2020,mahan_many-particle_2000}

\begin{acknowledgments}
	The authors are supported by the National Science Foundation of China under Grant No.~12174005, the National Key R\&D Program of China under Grand Nos.~2018YFA0305603 and 2021YFA1401900. The computational resources were provided by the High-performance Computing Platform of Peking University.
\end{acknowledgments}

\bibliographystyle{apsrev4-2}
\bibliography{Reference}

\end{document}


	
	\title{Supplemental Materials:\\ Enhancing superconducting transition temperature of lanthanum superhydride by increasing hydrogen vacancy concentration}
	
	
	\author{Haoran Chen}
	\email{chenhr1115@pku.edu.cn}
	\affiliation{International Center for Quantum Materials, Peking University, Beijing 100871, China}
	
	\author{Hui Wang}
	\affiliation{Key Laboratory for Photonic and Electronic Bandgap Materials (Ministry of Education), School of Physics and Electronic Engineering, Harbin Normal University, Harbin 150025, China}
	
	\author{Junren Shi}
	\affiliation{International Center for Quantum Materials, Peking University, Beijing 100871, China}
	\affiliation{Collaborative Innovation Center of Quantum Matter, Beijing 100871, China}
	
	
	
	\maketitle
	
	\setcounter{equation}{0}
	\setcounter{figure}{0}
	\setcounter{table}{0}
	\setcounter{page}{1}
	\makeatletter
	\renewcommand{\theequation}{S\arabic{equation}}
	\renewcommand{\thefigure}{S\arabic{figure}}
	\renewcommand{\thetable}{S\arabic{table}}
	\renewcommand{\bibnumfmt}[1]{[S#1]}
	\renewcommand{\citenumfont}[1]{S#1}
	\renewcommand{\thesection}{S\Roman{section}}

	\section{Numerical details}
	\subsection{Density functional theory calculations}
	All density functional theory (DFT) calculations are performed using the CPU and GPU version of the Vienna ab initio Simulation Package (VASP) code~\cite{kresse_efficient_1996,kresse_ultrasoft_1999}. The projector-augmented wave (PAW) method~\cite{blochl_projector_1994,kresse_ultrasoft_1999} is used to describe the ion-electron interaction, and the Perdew-Burke-Ernzerhof (PBE) functional~\cite{perdew_generalized_1996} is used to describe the exchange-correlation effect.
	
	\subsection{Machine-learning force field and path-integral molecular dynamics}
	Path-integral molecular dynamics (PIMD) simulations are performed using the  machine-learning force field (MLFF) technique. The MLFF is trained on the fly in PIMD simulations using the modified version of VASP previously developed in Ref.~\cite{chen_coexistence_2024}, which combines the on-the-fly-machine-learning built-in module of VASP~\cite{jinnouchi_--fly_2019,jinnouchi_phase_2019} and path-integral molecular dynamics (PIMD)~\cite{ceriotti_efficient_2010}
	
	For ``samples'' of each vacancy concentration, we train a corresponding MLFF.
	Training is performed in cubic supercells of volume 1097.70~\AA$^3$, 1062.39~\AA$^3$ and 1014.35~\AA$^3$, all containing 32 formulas of LaH$_{10-\delta}$.
	The supercells corresponds to a pressure \textit{close to} 137~GPa, 150~GPa and 176~GPa, respectively.
	(Path-integral) Langevin thermostat is used to control temperature in simulations~\cite{ceriotti_efficient_2010}. The friction coefficients of centroid mode are set to [1 1 1]~ps${}^{-1}$ and a time step of 1~fs is used.
	During the simulation, if a structure is judged to be outside the previous training sets, a DFT calculation is performed.
	In DFT calculations, we use an energy cutoff of 450~eV for plane waves to expand electron wavefunctions, and a $3\times3\times3$ $\Gamma$-centered $k$-point grid to sample the Brillouin zone of the supercell.
	
	The training is performed in three steps.
	First, an on-the-fly MD training is performed. The temperature is raised from 100~K to 400~K in a 20-ps simulation. Second, quantum effects are taken into account by performing a PIMD simulation with bead number (Trotter number) $N_b=8$. A 20-ps simulation is performed while raising the temperature from 100~K to 400~K. 
	Finally, we keep the temperature at 400~K, and perform another 20-ps simulation.
	In these steps, the radial and angular descriptors are constructed using parameters ML\_RCUT1=8.0, ML\_RCUT2=5.0, ML\_MBR1=14, ML\_MBR2=10 and ML\_SION=0.4. ML\_SCLC\_CTIFOR=0.6 is used to select local configurations.
	Typically, 550\,-\,650 structures and 180\,-\,200 (1400\,-\,1700) local configurations for La (H) are chosen for training, and the root-mean-square error (RMSE) of forces reaches 0.075\,-\,0.095~eV/\AA.
	
\begin{figure}
		\includegraphics[width=8.6cm]{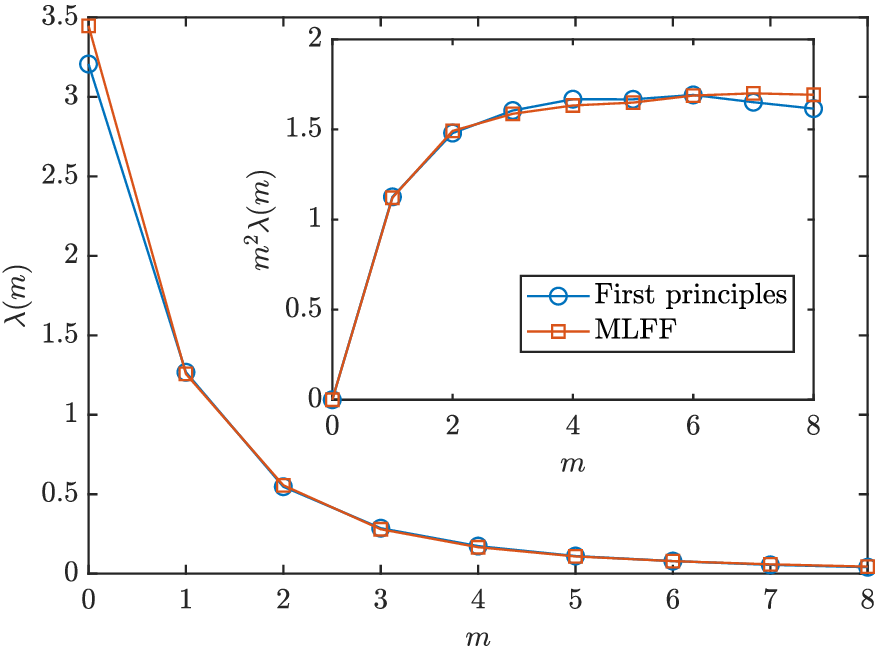}
		\caption{\label{fig:lam} EPC parameters $\lambda(m)$ from DFT-based and MLFF-based PIMD, respectively. The asymptotic behavior of $m^2\lambda(m)$ is also shown in the inset for comparison.}
	\end{figure}

	To test the force fields generated in this way, we use LaH$_{9.75}$ at 153~GPa as an example and perform calculations in a smaller supercell containing 4 formulas. 
	We first equilibrate the system using a DFT-PIMD simulation of 3~ps, following which both DFT-based and MLFF-based PIMD are performed for 3~ps.
	The simulations are performed at 240~K with bead number $N_b=16$.
	In Fig.~\ref{fig:lam}, we compare EPC parameters obtained in the two simulations, respectively. Satisfactory agreements are obtained.
	
%
	
	\subsection{Effective interaction and EPC parameters}
	Using the MLFF generated above, we simulate LaH$_{10-\delta}$ using PIMD at 240~K, with bead number $N_b$=16, time step 0.5~fs and an overall simulation time of 3~ps after equilibration.
	
	The ion configurations are uniformly sampled with a spacing of 40 time steps for subsequent stochastic path-integral approach (SPIA) calculations. DFT calculations are performed for these configurations. An energy cutoff of 350~eV for plane waves is used to expand electron wavefunctions, and a $2\times2\times2$ $\Gamma$-centered $k$-point grid to sample the Brillouin zone of the supercell.
	The converged local densities are then used as inputs of our MATLAB implementation of SPIA~\cite{chen_first-principles_2021,chen_stochastic_2022,chen_coexistence_2024}.
	The effective electron-electron interaction $\hat{W}$ are then calculated on the irreducible wedge of a $6\times6\times6$ $\bm{k}$-point grid. 
	The Fermi-surface-averaged EPC parameters $\lambda(m)$ are then determined by summing $\hat{W}$ on the Fermi surface~\cite{chen_stochastic_2022}, i.e.,
	\begin{eqnarray}\label{lambda_avg}
		\lambda(j-j')=-\frac{1}{N(\varepsilon_F)}\sum_{n\bm{k},n'\bm{k'}}
		W_{n\bm{k},n'\bm{k'}}(j-j')
		\delta(\varepsilon_{n\bm{k}}-\varepsilon_F)\delta(\varepsilon_{n'\bm{k'}}-\varepsilon_F),
	\end{eqnarray}
	where the index $j$ denotes the Fermion Matsubara frequency $\nu_j=(2j+1)\pi/\beta$, with $\beta=k_B T$.
	The delta function $\delta(\varepsilon_{n\bm{k}}-\varepsilon_F)$ is replaced with a Lorentzian function
	\begin{eqnarray}
		\delta(\varepsilon_{n\bm{k}}-\varepsilon_F)\rightarrow \frac{1}{\pi}\frac{\gamma}{(\varepsilon_{n\bm{k}}-\varepsilon_F)^2+\gamma^2}
	\end{eqnarray}
	in our calculation to approximate the imaginary part of a Green's function~\cite{liu_superconducting_2020}. The half width is set to $\gamma=0.16$~eV.

	\section{Pressure dependence of density of states}
	\begin{figure}
		\includegraphics[width=11.2cm]{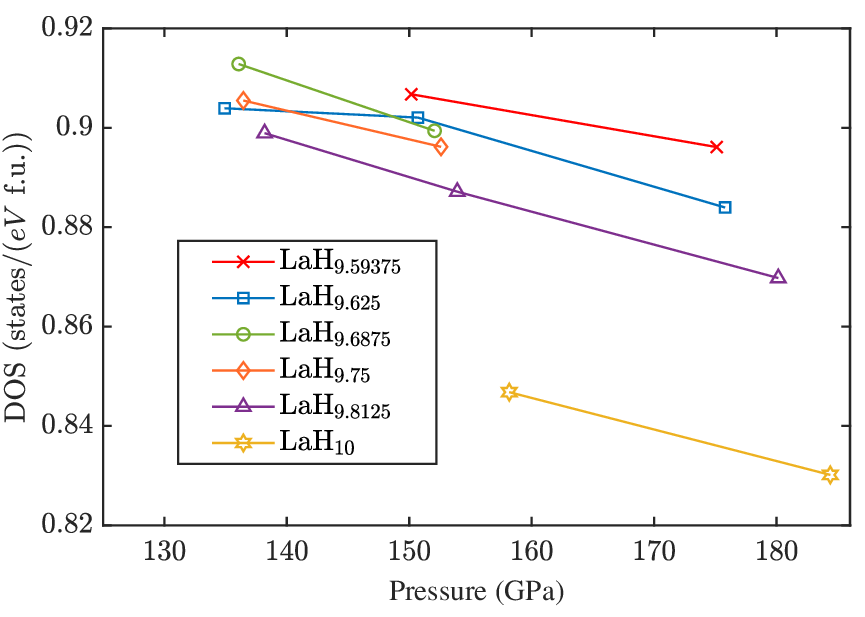}
		\caption{\label{fig:band}
			Pressure dependence of total DOS of LaH${_{10-\delta}}$with different vacancy-concentrations.}
	\end{figure}
	In Fig.~\ref{fig:band}~(c), we show the dependence of total density of states (DOS) at Fermi level on pressure, for LaH$_{10-\delta}$ of different vacancy concentrations. 
	The DOS is calculated by averaging over all PIMD configurations.
	It can be seen that, in all cases, DOS decreases when pressure increases.

	\section{Pressure dependence of $T_c$}
\begin{figure}[h]
		\includegraphics[width=11.2cm]{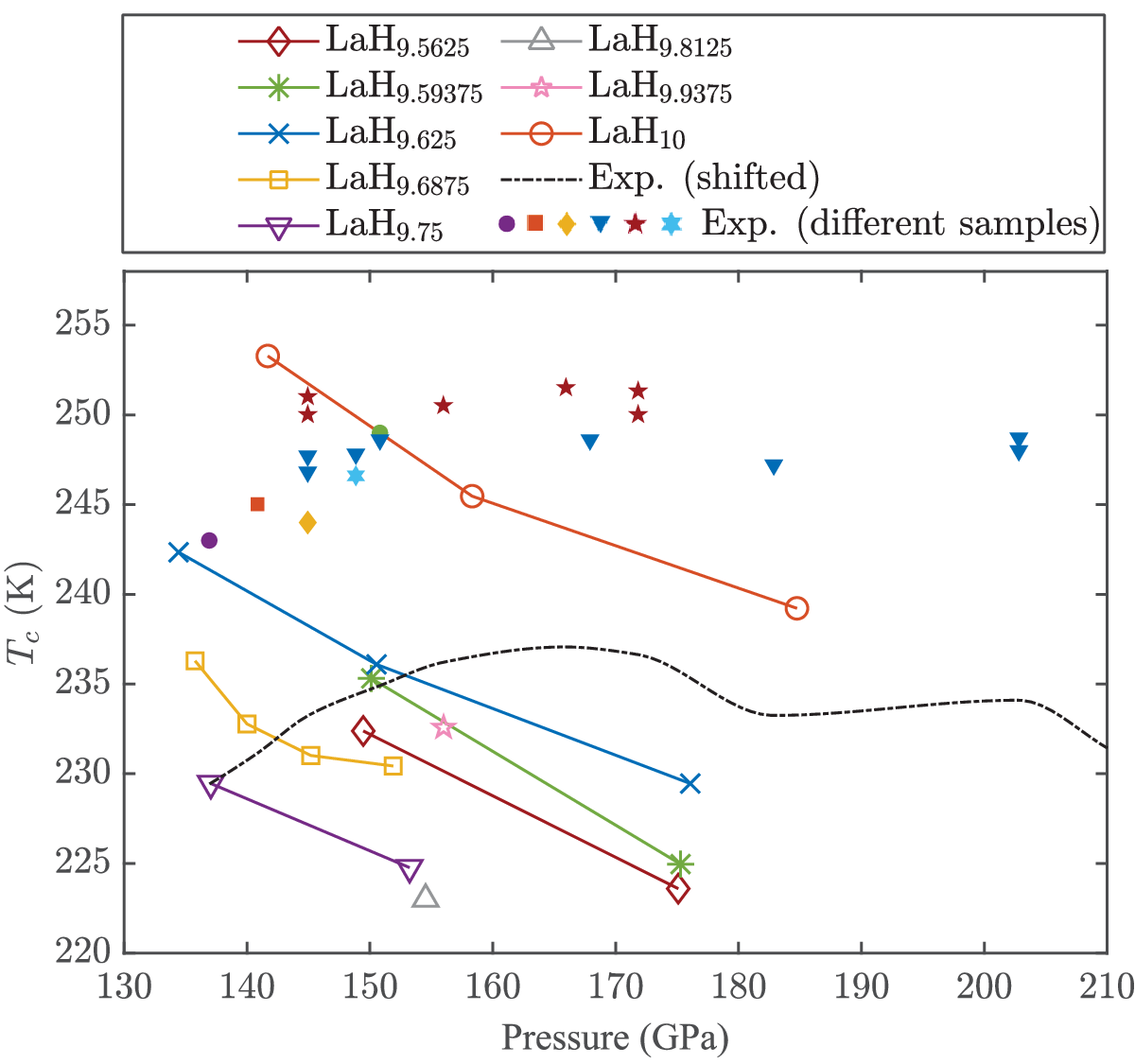}
		\caption{\label{fig:Tc}
			Pressure dependence of $T_c$ of LaH${_{10-\delta}}$ with different vacancy concentrations. 
			Experimental measurements (original and that downshifted by 13~K) from Ref.~\cite{drozdov_superconductivity_2019} are shown for comparison.}
	\end{figure}

	In this section, we show results about the pressure dependence of $T_c$ in LaH$_{10-\delta}$ of different $\delta$'s. As shown in Fig.~\ref{fig:Tc}, $T_c$ generally decreases upon compression for a fixed $\delta$.
	
	\section{Contribution from diffusion}
\begin{figure}
		\includegraphics[width=14.2cm]{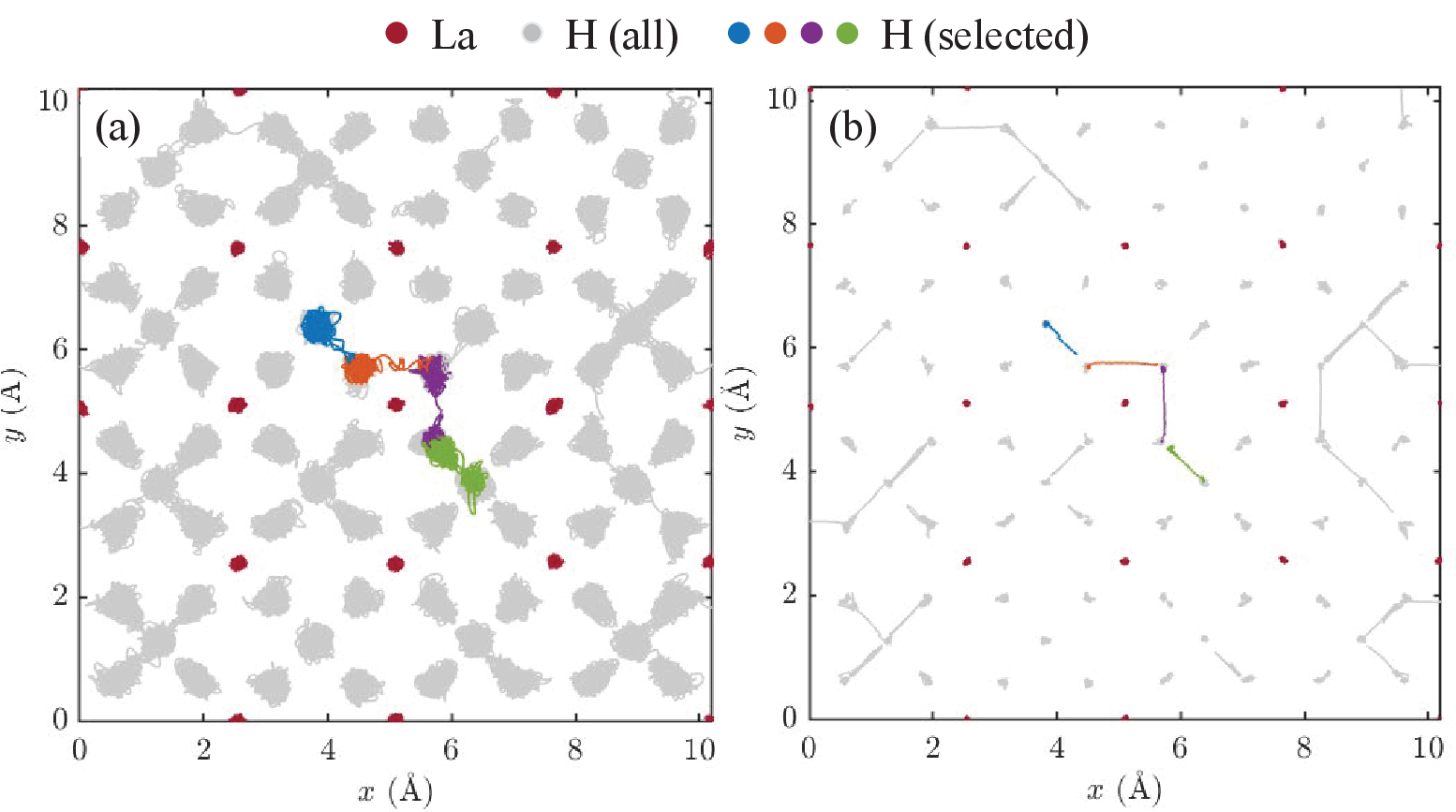}
		\caption{\label{fig:dif} 
			(a) [001] view of trajectories of the centroid mode of all hydrogen (grey) and lanthanum (red) ions in  LaH$_{9.625}$ under 151~GPa from PIMD simulations at 240~K. 
			Trajectories of four selected hydrogen ions are highlighted with different colors.
			(b) Same as (a), while the trajectories are smoothed using a moving-average filter with 250-fs window length.
		}
	\end{figure}

	In this section, we sketch how we describe the contribution from diffusion.
	Following Ref.~\cite{chen_coexistence_2024}, we assume that 	the trajectories of ions can be expressed as the summation of diffusion paths and local vibrations around the path:
	\begin{eqnarray}
		\bm{R}(t,\tau)=\bm{R}_{\mathrm{dif}}(t,\tau)+\Delta\bm{R}_{\mathrm{vib}}(t,\tau).
	\end{eqnarray}
	Since local vibrations are much faster than the diffusion,
	the diffusion path $\bm{R}_{\mathrm{dif}}(t,\tau)$ can be obtained by performing a moving average of ion trajectories (see Fig.~\ref{fig:dif}). The contribution from the averaged path to EPC parameters ($\lambda^{\mathrm{avg}}(m)$) is then calculated using SPIA.
	
	\begin{figure}[!h]
		\includegraphics[width=8.6cm]{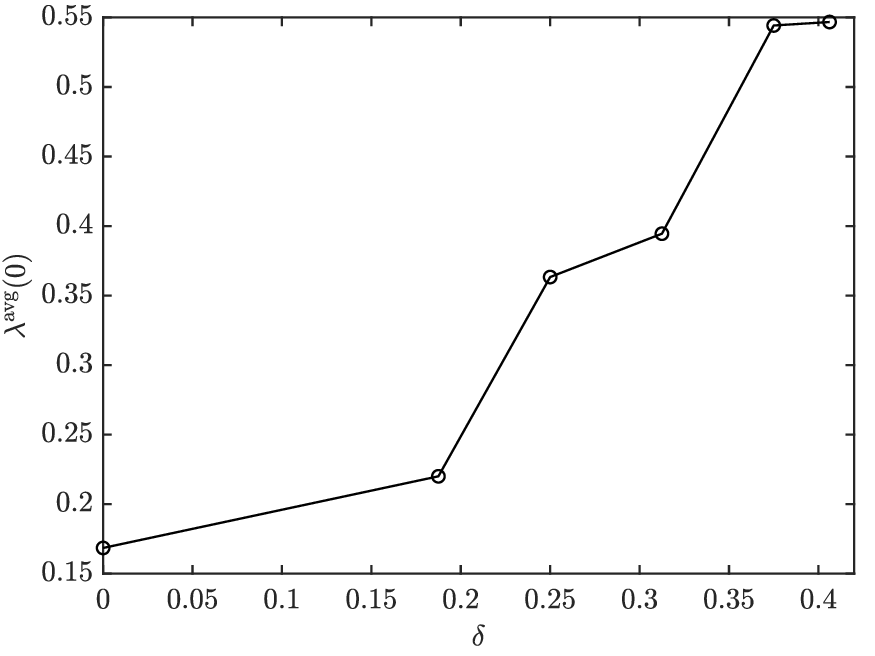}
		\caption{\label{fig:lam0} Static EPC parameter $\lambda^{\mathrm{avg}}(0)$ calculated from the moving-averaged paths of LaH$_{10-\delta}$ of different $\delta$'s.}
	\end{figure}
	
	It should be noted that, by performing moving average over ion trajectories, some low-frequency vibrations are inevitably included. This can be seen by performing average over the trajectories of LaH$_{10}$, in which no diffusion exists. As shown in Fig.~\ref{fig:lam0}, the slow vibrations also contribute a small $\lambda(0)$. 
	We assume this part of contribution is approximately unchanged when hydrogen vacancies are introduced, and define the diffusion EPC parameter as $\lambda^{\mathrm{dif}}_{\mathrm{LaH}_{10-\delta}}(0)=\lambda^{\mathrm{avg}}_{\mathrm{LaH}_{10-\delta}}(0)-\lambda^{\mathrm{avg}}_{\mathrm{LaH}_{10}}(0)$.
	
	It can also be seen that the contribution from diffusion is enhanced when we introduce more vacancies in the system.
	
	\section{Tests of convergence}
	In this section, we test the convergence of our calculations with respect to the sampling time step, simulation time and k-mesh density when calculating Green's functions and effective interactions. Tests are performed for LaH$_{9.625}$ at 151~GPa.
	
\begin{figure}
		\includegraphics[width=16.2cm]{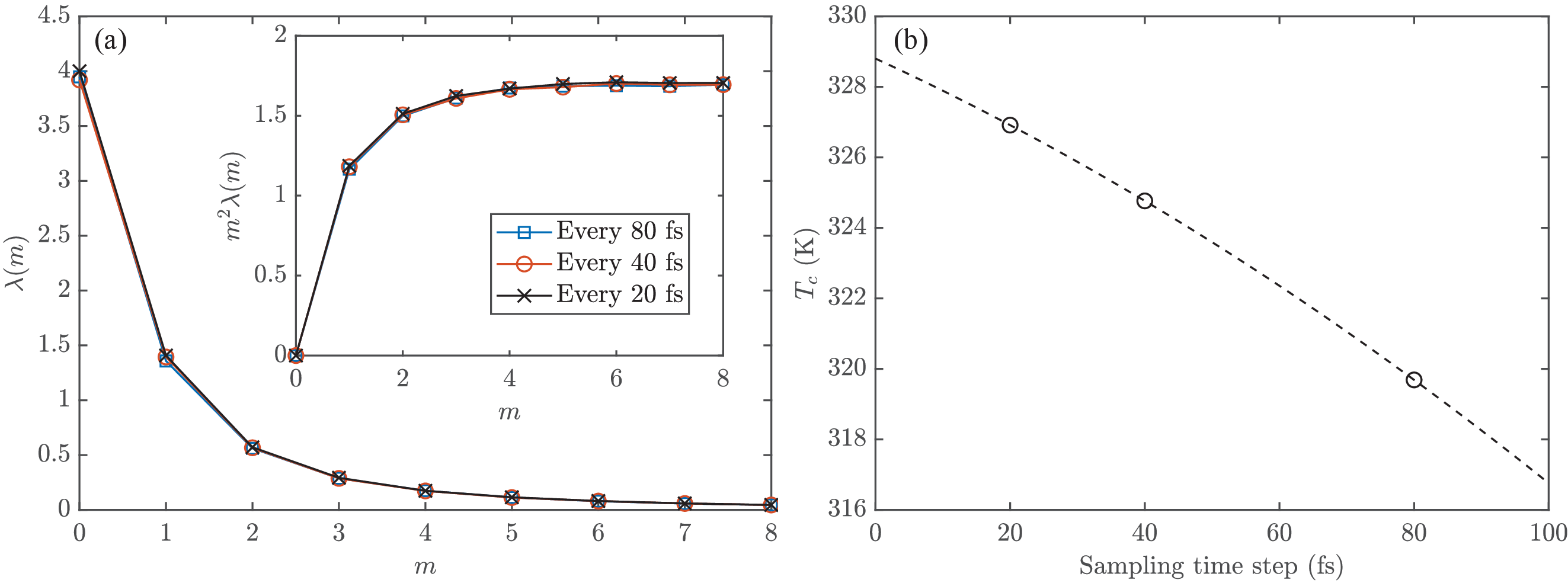}
		\caption{\label{fig:conv_dt} (a) EPC parameters $\lambda(m)$ and (b) predicted $T_c$ with respect to sampling time steps. The asymptotic behavior of $m^2\lambda(m)$ is also shown in the inset of (a) for comparison. A $2\times2\times2$ k-mesh of the supercell is used to calculate the parameters.}
	\end{figure}

	First, we test the sampling time step. 
	In Fig.~\ref{fig:conv_dt}, we show EPC parameters $\lambda(m)$ and $T_c$ calculated with configurations sampled every 20~fs, 40~fs and 80~fs. We see that a sampling step 20 fs is well converged.
	
\begin{figure}
		\includegraphics[width=16.2cm]{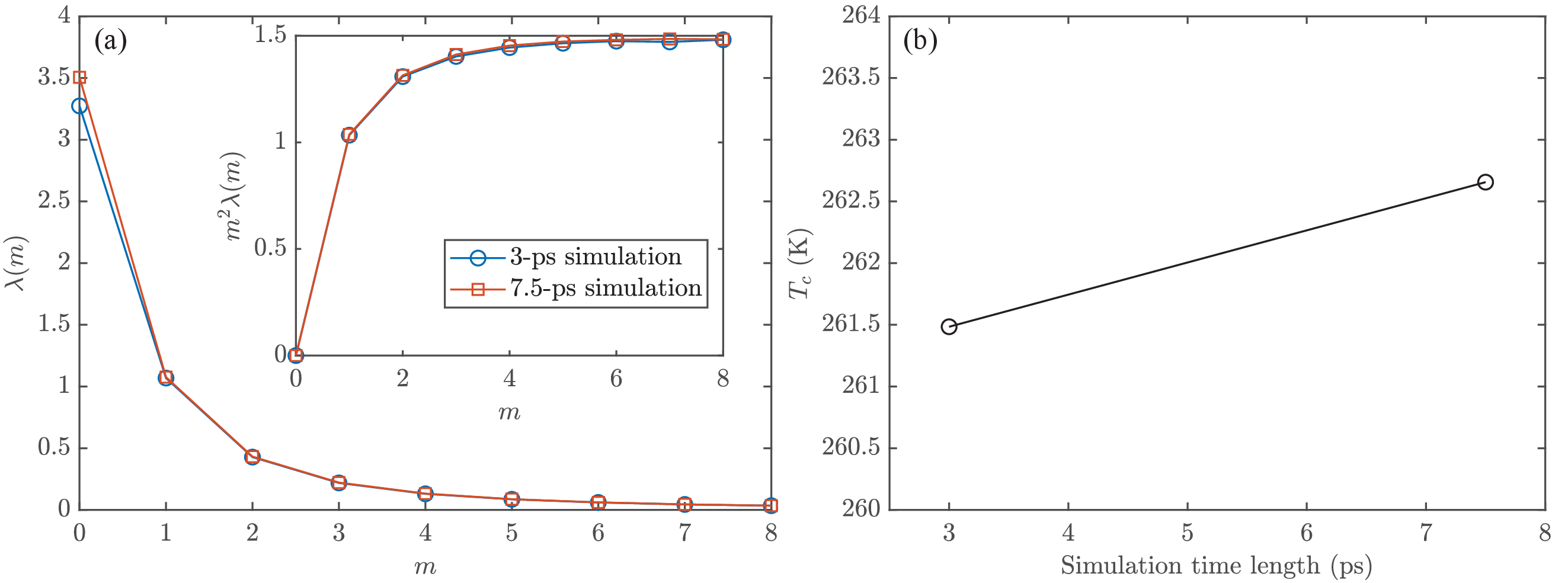}
		\caption{\label{fig:conv_time} (a) EPC parameters $\lambda(m)$ and (b) predicted $T_c$ with respect to simulation time lengths. The asymptotic behavior of $m^2\lambda(m)$ is also shown in the inset of (a) for comparison. A $5\times5\times5$ k-mesh of the supercell is used to calculate the parameters. Configurations are sampled every 20~fs from a 3-ps PIMD simulation.}
	\end{figure}

	Second, we test convergence regarding the simulation time length. As shown in Fig.~\ref{fig:conv_time}, longer simulation of 7.5~ps yields almost the same result as that from a 3-ps simulation.
	
\begin{figure}
		\includegraphics[width=16.2cm]{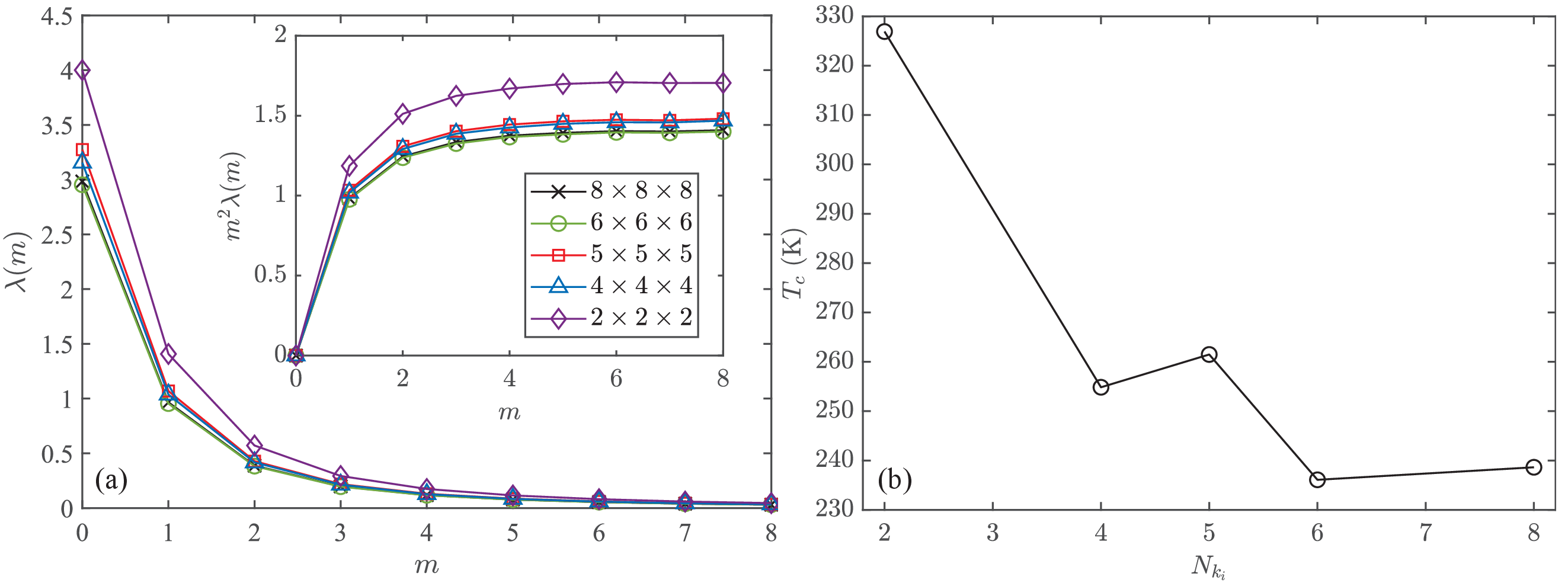}
		\caption{\label{fig:conv_k} (a) EPC parameters $\lambda(m)$ and (b) predicted $T_c$ with respect to the density . The asymptotic behavior of $m^2\lambda(m)$ is also shown in the inset of (a) for comparison. Configurations are sampled every 20~fs from a 3-ps PIMD simulation.}
	\end{figure}

	Finally, we test whether the k-mesh density is sufficient for determining $\lambda$ and $T_c$. As shown in Fig.~\ref{fig:conv_k}, the results from a $6\times6\times6$ k-mesh of the supercell is well converged.

	\FloatBarrier
	
	\bibliographystyle{apsrev4-2}
	\bibliography{Reference}